\documentclass[11pt]{article}

\fontfamily{times}
\usepackage{geometry}
\usepackage[compress,nospace]{cite}
\usepackage{float}
\usepackage{graphicx}
\usepackage{subfigure}
\usepackage{color}
\usepackage{amssymb}
\usepackage{amsmath}
\usepackage[final]{pdfpages}
\usepackage{amssymb}
\usepackage{amsfonts}
\usepackage{bm}
\usepackage{upgreek}
\usepackage{arydshln}
\usepackage{enumerate}
\usepackage{threeparttable}
\usepackage{caption}
\usepackage{multirow}
\usepackage{color}
\usepackage{xfrac}
\usepackage{array}

\newcommand{\templatefigures}[1]
{\noindent
\begin{minipage}{2cm}
\begin{center}
  \centering
	\vspace{-1cm}
  \includegraphics[scale=0.1]{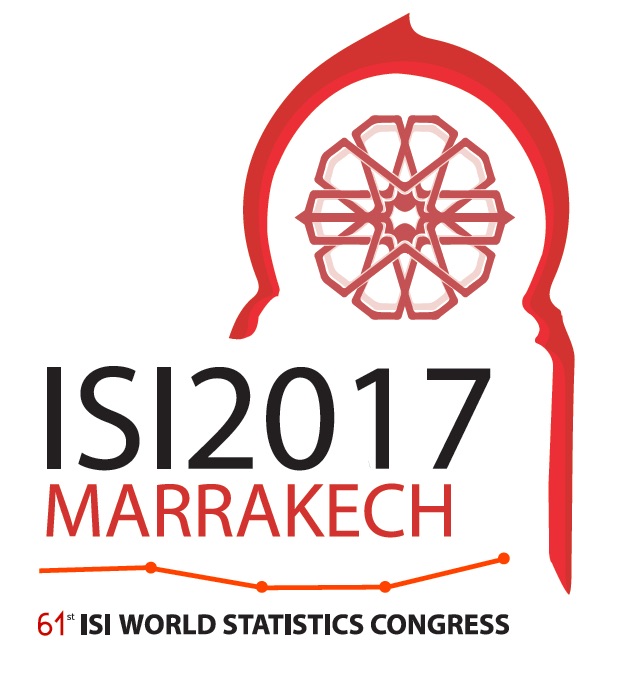}\\
\end{center}
\end{minipage}
\quad
\begin{minipage}{12cm}
\hspace*{6.8cm}
\end{minipage}
\quad
\begin{minipage}{2cm}
\begin{center}
\vspace{-0.9cm}
\includegraphics[scale=0.4]{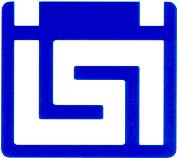}\\
\end{center}

\end{minipage}

\vskip0.2cm
}

\begin{document}
\templatefigures{}

\small{

\begin{center}
%The title should be centred and in bold letters. It should be informative but not too long (preferably no more than two lines).
\textbf{Asymptotic robustness of Kelly's GLRT and Adaptive Matched Filter detector under model misspecification}
\end{center}

\begin{center}
{Stefano Fortunati*}\\
{Dept. of Information Engineering, University of Pisa, Italy - stefano.fortunati@iet.unipi.it}\\ 
\vspace{0.5cm}

{Maria S. Greco}\\
{Dept. of Information Engineering, University of Pisa, Italy - m.greco@iet.unipi.it}\\
\vspace{0.5cm}

{Fulvio Gini}\\
{Dept. of Information Engineering, University of Pisa, Italy - f.gini@ing.unipi.it}\\

\end{center}

\begin{center}
{\bf Abstract}
\end{center}

\setlength{\parindent}{0pt}

A fundamental assumption underling any Hypothesis Testing (HT) problem is that the available data follow the parametric model assumed to derive the test statistic. Nevertheless, a perfect match between the true and the assumed data models cannot be achieved in many practical applications. In all these cases, it is advisable to use a robust decision test, i.e. a test whose statistic preserves (at least asymptotically) the same probability density function (pdf) for a suitable set of possible input data models under the null hypothesis. Building upon the seminal work of Kent (1982), in this paper we investigate the impact of the model mismatch in a recurring HT problem in radar signal processing applications: testing the mean of a set of Complex Elliptically Symmetric (CES) distributed random vectors under a possible misspecified, Gaussian data model. In particular, by using this general misspecified framework, a new look to two popular detectors, the Kelly's Generalized Likelihood Ration Test (GLRT) and the Adaptive Matched Filter (AMF), is provided and their robustness properties investigated. \\

}\\

\setlength{\parindent}{0pt}

{\bf 1. Introduction}

Testing between two (or more) statistical hypotheses represents a key task in many practical applications. A classical approach used to discriminate between two hypotheses, say $H_0$, the null hypothesis, and $H_1$, is to derive the Likelihood Ratio (LR) test from an assumed statistical model of the available dataset. The underlying assumption is that the acquired data actually follow the parametric model assumed to derive the test statistic. If this assumption is satisfied, then, \textit{asymptotically}, the LR statistic converges in distribution to a chi-square random variable with a suitable number of degrees of freedom (see e.g. Kay (1998)). In realistic scenarios however, the model assumed to derive the LR test and the true data model may differ to some extent. As a consequence of this possible model misspecification, the asymptotic distribution of the LR test statistic could no longer be a chi-square distribution. In his seminal work, Kent (1982) showed that, as the number of available data goes to infinity, the LR statistic under model misspecification converges in distribution to a weighted sum of independent chi-square random variables where the weights depends on the actual true and the assumed models. Starting from this result, Kent derived a criterion to select suitable parametric models for the acquired data for which the LR statistic is asymptotically robust, i.e. it is asymptotically chi-squared distributed irrespective to the actual true data model, provided that it belongs to a suitable class of distributions. Finally, another important contribution of Kent's paper is in the derivation of alternative, asymptotically chi-square distributed, robust tests. The aim of this paper is to investigate a possible applications of the main findings of Kent's paper to a subclass of HT problems that are frequently encounter in radar applications. In particular, we specialize Kent's results to the problem of testing the mean value of a set of Complex Elliptically Symmetric (CES) distributed random vectors (see e.g. Ollila et al. (2012)) when the assumed parametric model is a Complex Gaussian distribution. Clearly, the resulting HT problem is a \textit{composite} one since, to fully characterize the assumed Gaussian model, we have to rely on the estimate of its mean value (the parameter of interest) and of its covariance matrix that is generally referred to as a \textit{nuisance} parameter. The Generalized LR test (GLRT) and the Wald test are derived and their robustness property discussed for the particular model misspecification at hand. Finally, we show that the classical radar detection problem is a particular, single-sample, instance of such general HT problem and, as a consequence, two popular detectors, i.e. Kelly's GLRT (Kelly (1986)) and the AMF (Robey at al. (1992)), can be derived as special case of the (multiple-samples) GLRT and Wald test. Unfortunately, due to the non-asymptotic nature of the radar detection problem, the robustness considerations cannot be applied to Kelly's GLRT and the AMF.  \\

{\bf 2. A particular hypothesis testing problem under model misspecification}

In this paper, we consider the classical HT problem in radar applications of testing the absence (the $H_0$ hypothesis) or the presence (the $H_1$ hypothesis) of a known complex signal vector $\mathbf{v}$ in received data vectors $\mathbf{x}_m = \alpha\mathbf{v}+\mathbf{c}_m \in \mathbb{C}^{N}$ where $\mathbf{c}_m$ represents an unobserved noise random vector and $\alpha \in \mathbb{C}$ is an unknown, deterministic, signal parameter. Using the terminology frequently encountered in the radar literature, this detection problem can be recast as follow. Let $\mathbf{x}=\{\mathbf{x}_m\}_{m=1}^{M}$ be a set of $M$ independent observations vectors. We assume that the dataset $\mathbf{x}$ can be partitioned in: 
\begin{itemize}
	\item the primary dataset $\mathbf{x}^1=\{\mathbf{x}_m^1\}_{m=1}^{M_1}$, with $\mathbf{x}_m^1 = \alpha_1\mathbf{v}+\mathbf{c}_m$ and $\alpha_1=\alpha$,
	\item the secondary dataset $\mathbf{x}^0=\{\mathbf{x}_m^0\}_{m=1}^{M_0}$, with $\mathbf{x}_m^0 = \alpha_0\mathbf{v}+\mathbf{c}_m=\mathbf{c}_m$ and $\alpha_0=0$.
\end{itemize}
This data model can be used in radar scenarios involving distributed targets, i.e. objects that are spread over more than one range cells or in that applications in which the target echoes remains constant over a certain number of radar scans. Through this paper, we assume that each primary and secondary data vector $\mathbf{x}_{m}^i$, with $i=0,1$, is sampled from a CES distribution, i.e. $\mathbf{x}_{m}^i\sim CES_{N}(\alpha_{i},\boldsymbol{\Sigma},g), i=0,1$, whose pdf is given by:
\begin{equation}
\label{pdf_true}
p_{X_i}(\mathbf{x}_m) \triangleq p_{X_i}(\mathbf{x}_m;\alpha_i,\boldsymbol{\Sigma }) = c_{N,g}|\boldsymbol{\Sigma }|^{-1}g((\mathbf{x}_{m}-\alpha_i\mathbf{v})^{H}
\boldsymbol{\Sigma }^{-1}(\mathbf{x}_{m}-\alpha_i\mathbf{v})), i=0,1,
\end{equation}
where $c_{N,g}$ is a normalizing constant, $g(t):\mathbb{R}^{+} \rightarrow \mathbb{R}$ is the \textit{density generator}, $E_{p_{X_i}}\{\mathbf{x}_m\}=\alpha_i\mathbf{v}$ is the mean value and $\boldsymbol{\Sigma}$ is the full rank, positive definite Hermitian scatter/covariance matrix. Moreover, in order to deal with \textit{real}, unknown parameters, we define the vector $\boldsymbol{\eta}_i \triangleq [\mathrm{Re}(\alpha_i),\mathrm{Im}(\alpha_i)]^T$. Clearly, this detection problem can be formalized as:
\begin{equation}
\label{Model_selection_base}
H_0: p_{H_0}(\mathbf{x}) = \prod\nolimits_{m=1}^{M_1+M_0}p_{X_0}(\mathbf{x}_m)\quad \mathrm{vs} \quad H_1: p_{H_1}(\mathbf{x}) = \prod\nolimits_{m=1}^{M_1}p_{X_1}(\mathbf{x}_{m}^1)\prod\nolimits_{m=1}^{M}p_{X_0}(\mathbf{x}_m^0),
\end{equation}
where the marginal pdfs  $p_{X_i}(\mathbf{x}_m), i=0,1$ have the functional form in \eqref{pdf_true}. In order to discriminate $H_1$ against $H_0$ in the HT problem in \eqref{Model_selection_base}, we could rely on the Generalized LR Test (GLRT) statistic defined as:
\begin{equation}
\label{LR_stat_true}
\Lambda_{GLRT}(\mathbf{x}) \triangleq 2 \ln \left( \frac{\max_{\boldsymbol{\eta},\boldsymbol{\Sigma} }{ p_{H_1}(\mathbf{x};\boldsymbol{\eta},\boldsymbol{\Sigma})}}
{\max_{\boldsymbol{\Sigma}} p_{H_0}(\mathbf{x};\boldsymbol{0},\boldsymbol{\Sigma})} \right), \quad \boldsymbol{\eta} \triangleq [\mathrm{Re}(\alpha),\mathrm{Im}(\alpha)]^T.
\end{equation}
It is well-known that under the null hypothesis $H_0$, the GLRT decision statistic has an asymptotic central chi-square distribution, i.e. $
\Lambda_{GLRT}(\mathbf{x}|H_0) \sim\chi_p^2$, where $p=\dim(\boldsymbol{\eta})=2$, for $M_1,M_0\rightarrow\infty$ (Kay (1998)). Suppose now that we do not have any a-priori knowledge about the functional form of the density generator $g$ that characterizes the particular CES distribution of the acquired data. A direct consequence of this lack of knowledge is that the GLRT statistic in \eqref{LR_stat_true} cannot be evaluated. Note that this is a recurring cases in radar applications. Extensive experimental analyses have demonstrated that the statistical behaviour of the row radar data can be well described by the CES family, but in operative scenarios, we do not have the possibility to identify the best representative of the CES class that characterizes the actual stream of data. Consequently, motivated by the need to derive simple and real-time inference algorithms, many radar systems exploit the Complex Gaussian distribution as a surrogate data model. More formally, for both the primary and secondary data vectors we assume the following, possibly misspecified, Complex Gaussian distribution:
\begin{equation}
\label{Assumed_Gaussian}
f_{X_i}(\mathbf{x}_m,\alpha_i;\boldsymbol{\Phi}) = \pi^{-N}|\boldsymbol{\Phi }|^{-1}\exp((\mathbf{x}_{m}-\alpha_i\mathbf{v})^{H}
\boldsymbol{\Phi }^{-1}(\mathbf{x}_{m}-\alpha_i\mathbf{v})), i=0,1.
\end{equation}
Under this \textit{model misspecification}, the HT problem in \eqref{Model_selection_base} has to be recast as follow:
\begin{equation}
\label{Model_selection_Miss}
H_0: f_{H_0}(\mathbf{x}) = \prod\nolimits_{m=1}^{M_1+M_0}f_{X_0}(\mathbf{x}_m) \quad \mathrm{vs} \quad H_1: f_{H_1}(\mathbf{x}) = \prod\nolimits_{m=1}^{M_1}f_{X_1}(\mathbf{x}_{m}^1)\prod\nolimits_{m=1}^{M}f_{X_0}(\mathbf{x}_m^0).
\end{equation}
where the assumed joint pdfs $f_{H_i}(\mathbf{x};\alpha_i,\boldsymbol{\Phi})$ could differ from the true pdfs $p_{H_i}(\mathbf{x})$ for all the possible values of the signal parameter $\alpha_i$ and of the covariance matrix $\boldsymbol{\Phi }$, i.e $p_{H_i}(\mathbf{x})\neq f_{H_i}(\mathbf{x};\alpha_i,\boldsymbol{\Phi}), \forall \alpha_i,\boldsymbol{\Phi}$ under both the hypotheses $H_i, i=0,1$. A \textit{mismatched} GLRT (MGLRT) for the HT problem in \eqref{Model_selection_Miss} can be obtained as:
\begin{equation}
\label{MGLRT_stat}
\Lambda_{MGLRT}(\mathbf{x}) \triangleq 2 \ln \left( \frac{\max_{\boldsymbol{\eta},\boldsymbol{\Phi}}{ f_{H_1}(\mathbf{x};\boldsymbol{\eta},\boldsymbol{\Phi})}}
{\max_{\boldsymbol{\Phi}} f_{H_0}(\mathbf{x};\boldsymbol{0},\boldsymbol{\Phi})} \right), \quad \mathbf{x}_{m}^i\sim CES_{N}(\alpha_{i},\boldsymbol{\Sigma},g), i=0,1,
\end{equation}
where, as before, $\boldsymbol{\eta} \triangleq [\mathrm{Re}(\alpha),\mathrm{Im}(\alpha)]^T$. A closed form expression for $\Lambda_{MGLRT}(\mathbf{x})$ can be easily obtained following the same procedure discussed in Kelly (1986). We start by noticing that the assumed joint pdf of all the available data $\mathbf{x}$ can be expressed as:
\begin{equation}
\label{Assumed_Joint_Gaussian}
f_{H_i}(\mathbf{x}_m,\boldsymbol{\eta}_i;\boldsymbol{\Phi}) = \left( \pi^{-N}|\boldsymbol{\Phi }|^{-1}\exp[\mathrm{tr}(\boldsymbol{\Phi }^{-1}\mathbf{T}_i)]\right)^M, \quad M = M_1+M_0,
\end{equation}
where  
\begin{equation}
\label{T_0}
\mathbf{T}_0 \triangleq \frac{1}{M} \left( \mathbf{X}_1\mathbf{X}_1^H + \mathbf{S}_0\right)  =\frac{1}{M} \left( \sum\nolimits_{m=1}^{M_1}\mathbf{x}_m^1(\mathbf{x}_m^1)^H +  \sum\nolimits_{m=1}^{M_0}\mathbf{x}_m^0(\mathbf{x}_m^0)^H\right),
\end{equation}
\begin{equation}
\label{T_1} 
\mathbf{T}_1(\boldsymbol{\eta}) = \mathbf{T}_1(\mathrm{Re}(\alpha),\mathrm{Im}(\alpha)) \triangleq \frac{1}{M} \left( \tilde{\mathbf{X}}_1(\alpha)\tilde{\mathbf{X}}_1(\alpha)^H + \mathbf{S}_0\right)  =  \frac{1}{M}\left( \sum\nolimits_{m=1}^{M_1}(\mathbf{x}_m^1-\alpha\mathbf{v})(\mathbf{x}_m^1-\alpha\mathbf{v})^H + \mathbf{S}_0\right),
\end{equation}
\begin{equation}
\label{X_1} 
\mathbf{X}_1 = \left[ \mathbf{x}_1^1|\mathbf{x}_2^1|\cdots|\mathbf{x}_{M_1}^1\right] , \quad \tilde{\mathbf{X}}_1 = \left[ \mathbf{x}_1^1-\alpha\mathbf{v}|\mathbf{x}_2^1-\alpha\mathbf{v}|\cdots|\mathbf{x}_{M_1}^1-\alpha\mathbf{v}\right]. 
\end{equation}
In order to evaluate the MGLRT statistic in \eqref{MGLRT_stat}, $f_{H_i}(\mathbf{x}_m,\boldsymbol{\eta};\boldsymbol{\Phi})$ needs to be maximized over the signal parameter vector $\boldsymbol{\eta}=[\mathrm{Re}(\alpha),\mathrm{Im}(\alpha)]^T$ and over the covariance matrix $\boldsymbol{\Phi}$ for each of the two hypotheses. The maximizers, $\hat{\boldsymbol{\eta}}_i$ and $\hat{\boldsymbol{\Phi}}_i$, are, by definition, the Mismatched Maximum Likelihood (MML) estimates (see Huber (1967) and White (1982)) that have to be substituted in \eqref{MGLRT_stat} instead of the corresponding unknown true values. Following Kelly (1986), it is easy to verify that the MML estimator of the covariance matrix $\boldsymbol{\Phi}_i, i=0,1$ under the $H_0$ and $H_1$ hypotheses, are simply given by $\hat{\boldsymbol{\Phi}}_0 = \mathbf{T}_0$ and $\hat{\boldsymbol{\Phi}}_1 = \mathbf{T}_1$. Consequently, $\Lambda_{MGLRT}(\mathbf{x})$ can be rewritten as:
\begin{equation}
\label{MGLRT_stat_step_1}
\Lambda_{MGLRT}(\mathbf{x}) = 2M \ln \left( \frac{|\mathbf{T}_0|}{\min_{\boldsymbol{\eta}}|\mathbf{T}_1(\boldsymbol{\eta})|} \right) = 2M \ln \left( \frac{|\mathbf{I}_{M_1} + \mathbf{X}_1^H \mathbf{S}_0^{-1}\mathbf{X}_1^H|}{|\mathbf{I}_{M_1} + \tilde{\mathbf{X}}_1(\hat{\boldsymbol{\eta}})^H \mathbf{S}_0^{-1}\tilde{\mathbf{X}}_1(\hat{\boldsymbol{\eta}})|} \right), 
\end{equation}    
where the last equality follows directly from the Sylvester's determinant identity and $\mathbf{I}_{M_1}$ indicates the identity matrix with dimension $M_1\times M_1$. Moreover, it can be shown that the MML estimate of $\hat{\boldsymbol{\eta}}_1$ is given by (Bandiera et al., (2007)):
\begin{equation}
\label{MML_alpha}
\hat{\boldsymbol{\eta}}=\left[ \mathrm{Re}(\hat{\alpha}),\mathrm{Im}(\hat{\alpha})\right] ^T \quad \mathrm{where} \quad \hat{\alpha} = \frac{1}{M_1}\sum_{m=1}^{M_1} \frac{\mathbf{v}^H\mathbf{S}_0^{-1}\mathbf{x}_m^1}{\mathbf{v}^H\mathbf{S}_0^{-1}\mathbf{v}}.
\end{equation}
It can be noted that, if $M_1=1$, i.e. if we have only one primary data vector, the GLRT in \eqref{MGLRT_stat_step_1} becomes the well-known Kelly's GLRT derived in Kelly (1986) as:
\begin{equation}
\label{GLRT_Kelly}
\Lambda_{MGLRT}(\mathbf{x}) \equiv \Lambda_{Kelly}(\mathbf{x})= -2(M_0+1) \ln \left( 1 - \frac{|\mathbf{v}^H\mathbf{S}_0^{-1}\mathbf{x}_1^1|^2}{(\mathbf{v}^H\mathbf{S}_0^{-1}\mathbf{v})[1+(\mathbf{x}_1^1)^H\mathbf{S}_0^{-1}\mathbf{x}_1^1]}\right). 
\end{equation} 
After having derived the GLRT under the above-mentioned model misspecification, it is of interest to investigate its asymptotic behaviour under the $H_0$ hypothesis. The knowledge of the asymptotic test distribution will allow us to set a threshold in order to achieve the desired level of significance. For this reason, it would be desirable that the asymptotic test distribution does not depend on the true but unknown actual CES distribution of the input data. Such a robustness property of the MGLRT in \eqref{MGLRT_stat_step_1} is the focus of the next Section. \\

{\bf 3. Asymptotic robustness of the $\Lambda_{MGLRT}(\mathbf{x})$ under $H_0$}

The main aim of this Section is to show that, under $H_0$ the MGLRT statistic in \eqref{MGLRT_stat_step_1} converges in distribution to a chi-square random variable irrespective of the actual CES distribution of the input data vectors. This result is of great practical importance since it implies that the level of significance does not depends on the actual CES input data model. This important fact can be proved by relying on the Theorem 3.1 in Kent (1982). As a preliminary result, some properties of the joint MML estimators of $\boldsymbol{\eta}$ and of the covariance matrix $\boldsymbol{\Phi }$ need to be discussed. For notation simplicity, in the following we define the vector $\boldsymbol{\mu} \triangleq \mathrm{vecs}(\boldsymbol{\Phi})$, where the vecs operator is the "symmetric" counterpart of the classical vec operator.\\ 

\setlength{\parindent}{0pt}

\textit{3.1 Convergence of the MML estimator under the $H_0$ hypothesis}

In this subsection, we analyse the convergence and the asymptotic distribution of the joint MML estimator of $\boldsymbol{\eta}$ and of $\boldsymbol{\mu} \triangleq \mathrm{vecs}(\boldsymbol{\Phi})$ when we assume the Complex Gaussian model $f_{H_1}(\mathbf{x};\boldsymbol{\eta},\boldsymbol{\mu})$, while the available data are distributed according to the true bur unknown CES pdf $p_{H_0}(\mathbf{x})$. Following the results of Huber (1967) and White (1982), we have that the joint MML estimator $\hat{\boldsymbol{\theta}} \triangleq [\hat{\boldsymbol{\eta}}^T,\hat{\boldsymbol{\mu}}^T]^T$  converges \textit{almost surely} (\textit{a.s.}) to the point that minimizes the Kullback-Leibler divergence (KLD) between the true pdf $p_{H_0}(\mathbf{x})$ the assumed pdf $f_{H_1}(\mathbf{x};\boldsymbol{\theta})$, i.e. $D(p_{H_0}(\mathbf{x})\lVert f_{H_1}(\mathbf{x};\boldsymbol{\theta})) \triangleq E_p \{ \ln ( p_{H_0}(\mathbf{x})/f_{H_1}(\mathbf{x};\boldsymbol{\theta})) \}$. In particular, we have that
\begin{equation}
\label{conv_MML_H1}
\hat{\boldsymbol{\theta}} \overset{a.s.}{\underset{M_1,M_0\rightarrow\infty}{\rightarrow}}\bar{\boldsymbol{\theta}}=[\bar{\boldsymbol{\eta}}^T,\bar{\boldsymbol{\mu}}^T]^T,
\end{equation}
where the so-called pseudo-true parameter vector $\bar{\boldsymbol{\theta}}$ is defined as:
\begin{equation}
\label{pt_def_H1}
\bar{\boldsymbol{\theta}} \triangleq {\mathrm{argmin}}_{\boldsymbol{\theta}}\left\lbrace  D(p_{H_0}(\mathbf{x})\lVert f_{H_1}(\mathbf{x};\boldsymbol{\theta}) \right\rbrace  =  \mathrm{argmin}_{\boldsymbol{\eta} ,\boldsymbol{\mu}} \left\lbrace  - E_{p_{H_0}} \left\lbrace   \ln f_{H_1}(\mathbf{x};\boldsymbol{\eta},\boldsymbol{\mu}) \right\rbrace    \right\rbrace. 
\end{equation}
%\begin{equation}
%\label{pt_def_H1}
%\begin{split}
%\bar{\boldsymbol{\theta}} & \triangleq \underset{\boldsymbol{\theta} \in \Theta^d}{\mathrm{argmax}} \left\lbrace  D(p_{H_0}(\mathbf{x})\lVert %f_{H_1}(\mathbf{x};\boldsymbol{\theta}) \right\rbrace  \\
%& =  \underset{\boldsymbol{\eta} \in \Omega^p,\boldsymbol{\mu} \in \Xi^q}{\mathrm{argmax}} \left\lbrace  -\sum\nolimits_{m=1}^{M_1} E_{p_{H_0}} %\left\lbrace   \ln f_{X_1}(\mathbf{x}_m^1;\boldsymbol{\eta},\boldsymbol{\mu}) \right\rbrace  -  \sum\nolimits_{m=1}^{M_0} E_{p_{H_0}} %\left\lbrace \ln f_{X_0}(\mathbf{x}_m^0;\mathbf{0};\boldsymbol{\mu}) \right\rbrace   \right\rbrace. 
%\end{split}
%\end{equation}
Moreover, $\hat{\boldsymbol{\theta}}$ is asymptotically unbiased (with respect to $\bar{\boldsymbol{\theta}}$) and normal distributed as:
\begin{equation}
\label{asym_MML_H1}
\left( \begin{array}{c}
\sqrt{M_1}(\hat{\boldsymbol{\eta}} - \bar{\boldsymbol{\eta}})\\
\sqrt{M_1+M_0} (\hat{\boldsymbol{\mu}} - \bar{\boldsymbol{\mu}})
\end{array}\right) 
{\underset{M_1,M_0\rightarrow\infty}{\sim}} \mathcal{N}\left( \mathbf{0}, \mathbf{C}(\bar{\boldsymbol{\theta}}) \right),
\end{equation}
where
\begin{equation}
\label{C_0_H1}
\mathbf{C}(\bar{\boldsymbol{\theta}}) \triangleq \mathbf{A}^{-1}(\bar{\boldsymbol{\theta}})\mathbf{B}(
\bar{\boldsymbol{\theta}})\mathbf{A}(\bar{\boldsymbol{\theta}})^{-1} = 
\left( \begin{array}{cc}
\mathbf{C}_{\boldsymbol{\eta}}(\bar{\boldsymbol{\theta}}) & \mathbf{C}_{\boldsymbol{\eta}\boldsymbol{\mu}}(\bar{\boldsymbol{\theta}})\\
\mathbf{C}_{\boldsymbol{\mu}\boldsymbol{\eta}}(\bar{\boldsymbol{\theta}}) & \mathbf{C}_{\boldsymbol{\mu}}(\bar{\boldsymbol{\theta}})
\end{array}\right),
\end{equation}

\begin{equation}
\label{A_0}
\mathbf{A}(\bar{\boldsymbol{\theta}}) = 
\left( \begin{array}{cc}
\mathbf{A}_{\boldsymbol{\eta}}(\bar{\boldsymbol{\theta}}) & \mathbf{A}_{\boldsymbol{\eta}\boldsymbol{\mu}}(\bar{\boldsymbol{\theta}})\\
\mathbf{A}_{\boldsymbol{\mu}\boldsymbol{\eta}}(\bar{\boldsymbol{\theta}}) & \mathbf{A}_{\boldsymbol{\mu}}(\bar{\boldsymbol{\theta}})
\end{array}\right) = 
E_{p_{H_0}} \left\{ \nabla^{T}_{\bar{\boldsymbol{\theta}}} \nabla_{\bar{\boldsymbol{\theta}}} \ln f_{H_1}(\mathbf{x};\bar{\boldsymbol{\theta}}), \right\},
\end{equation}
\begin{equation}
\label{B_0}
\mathbf{B}(\bar{\boldsymbol{\theta}})= 
\left( \begin{array}{cc}
\mathbf{B}_{\boldsymbol{\eta}}(\bar{\boldsymbol{\theta}}) & \mathbf{B}_{\boldsymbol{\eta}\boldsymbol{\mu}}(\bar{\boldsymbol{\theta}})\\
\mathbf{B}_{\boldsymbol{\mu}\boldsymbol{\eta}}(\bar{\boldsymbol{\theta}}) & \mathbf{B}_{\boldsymbol{\mu}}(\bar{\boldsymbol{\theta}})
\end{array}\right) =
E_{p_{H_0}} \left\{ \nabla_{\bar{\boldsymbol{\theta}}} \ln f_{H_1}(\mathbf{x};\bar{\boldsymbol{\theta}}) \nabla^{T}_{\bar{\boldsymbol{\theta}}} \ln f_{H_1}(\mathbf{x};\bar{\boldsymbol{\theta}}) \right\}.
\end{equation}
We start by evaluating the pseudo-true parameter vector in \eqref{pt_def_H1}. Under the standard regularity conditions, $\bar{\boldsymbol{\theta}}$ can be evaluated by solving the following system of non linear equations:
\begin{equation}
\label{system_pt}
\left\{ \begin{array}{c}
E_{p_{H_0}} \{\nabla_{\boldsymbol{\eta}} \ln f_{H_1}(\mathbf{x};\boldsymbol{\eta},\boldsymbol{\mu})\}|_{\boldsymbol{\eta}=\bar{\boldsymbol{\eta}}}=\boldsymbol{0}\\
E_{p_{H_0}} \{\nabla_{\boldsymbol{\mu}}\ln f_{H_1}(\mathbf{x};\boldsymbol{\eta},\boldsymbol{\mu})\}|_{\boldsymbol{\mu}=\bar{\boldsymbol{\mu}}}=\boldsymbol{0}\end{array} \right..
\end{equation}
It is immediate to verify that, by substituting \eqref{Assumed_Gaussian} in the first equation, we get:
\begin{equation}
\label{pt_0}
- \sum\nolimits_{m=1}^{M_1}E_{p_{H_0}} \{(\mathbf{x}_m^1-\alpha\mathbf{v})^H\}\boldsymbol{\Phi}^{-1}\mathbf{v} |_{\alpha=\bar{\alpha}}=0 \quad \Rightarrow \quad \bar{\boldsymbol{\eta}} \triangleq [\mathrm{Re}(\bar{\alpha}),\mathrm{Im}(\bar{\alpha})]^T = \boldsymbol{0}.
\end{equation}
Consequently, by substituting $\bar{\boldsymbol{\eta}} = \boldsymbol{0}$ in the second equation of \eqref{system_pt} and by exploiting the results shown in Fortunati at al. (2016), it can be proved that $\bar{\boldsymbol{\mu}}=\mathrm{vecs}(\boldsymbol{\Sigma})$, where $\boldsymbol{\Sigma}$ is the true covariance matrix of the actual pdf $p_{H_0}(\mathbf{x})$ in \eqref{pdf_true}. Then, we can say that the joint MML estimator converges to the true mean value, i.e. $\boldsymbol{0}$, and to the true covariance matrix $\boldsymbol{\Sigma}$, i.e.
\begin{equation}
\label{conv_MML_H1_result}
\hat{\boldsymbol{\theta}} \triangleq [\hat{\boldsymbol{\eta}}^T,\mathrm{vecs}(\hat{\boldsymbol{\Phi}})^T]^T = [\hat{\boldsymbol{\eta}}^T,\mathrm{vecs}(\mathbf{T}_1(\hat{\boldsymbol{\eta}}))^T]^T \overset{a.s.}{\underset{M_1,M_0\rightarrow\infty}{\rightarrow}}\bar{\boldsymbol{\theta}}=[\boldsymbol{0}^T,\mathrm{vecs}(\boldsymbol{\Sigma})^T]^T,
\end{equation}
and, consequently, $\hat{\boldsymbol{\theta}}$ is a \textit{consistent} estimator under the $H_0$ hypothesis. Moreover, through direct calculation (see also Kano et al. (1993)), irrespective of the particular CES input data distribution, we have that:
\begin{equation}
\label{A_0_CES}
\mathbf{A}_{\boldsymbol{\eta}\boldsymbol{\mu}}(\bar{\boldsymbol{\theta}}) =\mathbf{B}_{\boldsymbol{\eta}\boldsymbol{\mu}}(\bar{\boldsymbol{\theta}})= \boldsymbol{0}, \quad\mathbf{A}_{\boldsymbol{\eta}}(\bar{\boldsymbol{\theta}})=\mathbf{B}_{\boldsymbol{\eta}}(\bar{\boldsymbol{\theta}}) = 
2\left( \begin{array}{cc}
\mathbf{v}^H\boldsymbol{\Sigma}^{-1}\mathbf{v} & 0\\
0 & \mathbf{v}^H\boldsymbol{\Sigma}^{-1}\mathbf{v}
\end{array}\right).
\end{equation}
Having established these results on the MML estimator, we are now ready to prove the robustness property of the MGLRT in \eqref{MGLRT_stat_step_1} under the $H_0$ hypothesis.\\

\setlength{\parindent}{0pt}

\textit{3.2 Asymptotic distribution of $\Lambda_{MGLRT}(\mathbf{x})$ under $H_0$}

To derive the asymptotic distribution of the MGLRT in \eqref{MGLRT_stat_step_1} for the misspecified HT problem at hand, we can directly apply the Theorem 3.1 in Kent. It must be noted that a prerequisite for the applicability of Theorem 3.1 is that the MML estimator in \eqref{conv_MML_H1} is consistent, at least for the parameters of interest, under the $H_0$ hypothesis. In our case study, this assumption is satisfied as discussed in the previous Section (see \eqref{conv_MML_H1_result}). Then, following Kent's findings, we have that the MGLRT statistic $\Lambda_{MGLRT}(\mathbf{x}|H_0)$ in \eqref{MGLRT_stat_step_1} is asymptotically distributed as
\begin{equation}
\label{LR_mis_stat}
\Lambda_{MGLRT}(\mathbf{x}|H_0) \underset{M_1,M_0\rightarrow\infty}{\sim} \sum\nolimits_{i=1}^{p}\lambda_i v_i, \quad v_i \sim \chi_1^2 \quad \mathrm{and} \quad p=2,
\end{equation} 
where $\{\lambda_i\}_{i=1}^p$ are the eigenvalues of the matrix $\mathbf{H}(\bar{\boldsymbol{\theta}}) \triangleq \mathbf{P}(\bar{\boldsymbol{\theta}}) \mathbf{C}_{\boldsymbol{\eta}}(\bar{\boldsymbol{\theta}})$ with
\begin{equation}
\label{P_Matrix}
\mathbf{P}(\bar{\boldsymbol{\theta}}) \triangleq  \mathbf{A}_{\boldsymbol{\eta}}(\bar{\boldsymbol{\theta}})-\mathbf{A}_{\boldsymbol{\eta}\boldsymbol{\mu}}(\bar{\boldsymbol{\theta}})\mathbf{A}_{\boldsymbol{\mu}}(\bar{\boldsymbol{\theta}})^{-1}\mathbf{A}_{\boldsymbol{\mu}\boldsymbol{\eta}}(\bar{\boldsymbol{\theta}}).
\end{equation}
Moreover, $\Lambda_{MGLRT}(\mathbf{x}|H_0)$ is asymptotically equivalent to the misspecified Wald statistic
\begin{equation}
\label{Wald_test}
W(\mathbf{x}|H_0) \triangleq M_1 \hat{\boldsymbol{\eta}}^T \hat{\mathbf{P}}\hat{\boldsymbol{\eta}},
\end{equation}
where $\hat{\mathbf{P}}$ is any consistent estimate of the matrix $\mathbf{P}(\bar{\boldsymbol{\theta}})$. From \eqref{A_0_CES}, it is immediate to verify that $ \mathbf{P}(\bar{\boldsymbol{\theta}}) = \mathbf{A}_{\boldsymbol{\eta}}(\bar{\boldsymbol{\theta}})$ and $\mathbf{H}(\bar{\boldsymbol{\theta}})=\mathbf{I}_2$ and consequently:
\begin{equation}
\label{LR_mis_sim}
\Lambda_{MGLRT}(\mathbf{x}|H_0) \underset{M_1,M_0\rightarrow\infty}{\sim} \chi_2^2,\quad \mathbf{x}_{m}^i\sim CES_{N}(\mathbf{0},\boldsymbol{\Sigma},g), i=0,1.
\end{equation}
This result highlights the asymptotic robustness of the GLRT derived for the misspecified HT problem at hand. Specifically, the MGLRT in \eqref{MGLRT_stat_step_1} converges in distribution to a chi-square random variable irrespective of the true but possibly unknown CES data distribution. The most important consequence of this robustness property is in the fact that, asymptomatically, the level of significance of the test depends only on the chosen threshold or, to use the radar terminology, the MGLRT in \eqref{MGLRT_stat_step_1} is (asymptomatically) a \textit{Constant False Alarm Rate, (CFAR)} detector w.r.t. the actual CES data distribution. To conclude, we show now that, exactly as the celebrated Kelly's GLRT can be obtained as a particular (single-sample) instance of the general MGLRT in \eqref{MGLRT_stat_step_1}, the well-known AMF is the single-sample version of the misspecified Wald Test in \eqref{Wald_test}. We start by noticing that, from \eqref{A_0_CES}, a consistent estimate of the matrix $\mathbf{P}(\bar{\boldsymbol{\theta}})$ in \eqref{P_Matrix} can be obtained as:
\begin{equation}
\label{P_est}
\hat{\mathbf{P}}=\mathbf{A}_{\boldsymbol{\eta}}(\hat{\boldsymbol{\theta}})= 
2\left( \begin{array}{cc}
\mathbf{v}^H\mathbf{T}_1(\hat{\boldsymbol{\eta}})^{-1}\mathbf{v} & 0\\
0 & \mathbf{v}^H\mathbf{T}_1(\hat{\boldsymbol{\eta}})^{-1}\mathbf{v}
\end{array}\right).
\end{equation}
By substituting \eqref{MML_alpha} and \eqref{P_est} in \eqref{Wald_test}, we obtain an explicit expression of the misspecified Wald test for the HT problem at hand as:
\begin{equation}
\label{AMF}
W(\mathbf{x}) = 2 \frac{(\mathbf{v}^H\mathbf{T}_1(\hat{\boldsymbol{\eta}})^{-1}\mathbf{v})}{M_1(\mathbf{v}^H\mathbf{S}_0^{-1}\mathbf{v})^2}  \left| \sum\nolimits_{m=1}^{M_1} \mathbf{v}^H\mathbf{S}_0^{-1}\mathbf{x}_m^1\right| ^2.
\end{equation}
Finally, it is immediate to verify that, by using the matrix $M_0^{-1}\mathbf{S}_0$ as the consistent estimate of $\mathbf{\Sigma}$ under the $H_0$ hypothesis instead of $\mathbf{T}_1(\hat{\boldsymbol{\eta}})$, in the single-sample case ($M_1=1$), the misspecified Wald test simplified to the well-known AMF: 
\begin{equation}
\label{AMF_single_S}
W(\mathbf{x}) \equiv \Lambda_{AMF}(\mathbf{x}) = 2 M_0 \frac{\left| \mathbf{v}^H\mathbf{S}_0^{-1}\mathbf{x}_1^1\right| ^2} {\mathbf{v}^H\mathbf{S}_0^{-1}\mathbf{v}}.
\end{equation}
\\

{\bf 4. Conclusions}

Recently, the use of robust inference methods able to face with possible, and often unavoidable, data model misspecification is gaining increasing attention among the Signal Processing (SP) community. Even if a lot of valuable findings on this field are already available in the statistical literature, they still remain largely unrecognised by the majority of the SP practitioners. The aim of this paper was then to show that a fundamental result about the asymptotic robustness of the LR test in the presence of model misspecification provided by Kent (1982) can be successfully applied to the classical radar detection problem, i.e. the problem of discriminating between the presence or the absence of a target in a given area. Moreover, we showed that two well-known detectors, the Kelly's GLRT and the AMF, can be derived as single-sample instances of the more general framework proposed in Kent (1982). Of course, a lot of work still needs to be done. In particular, the detection algorithms discussed in this paper are only asymptotically robust.  However, the derivation of inference methods that satisfy the robustness property also in the finite sample regime and even in the single-sample case is of great practical importance and remains an open area for research.
\\

\setlength{\parindent}{0pt}

\textit{Acknowledgement}

The work of Stefano Fortunati has been supported by the Air Force Office of Scientific Research under award number FA9550-17-1-0065.\\

{\bf References}

Bandiera, F., Besson, O., Orlando, D., Ricci, G., Scharf L. L. (2007). GLRT-Based Direction Detectors in Homogeneous Noise and Subspace Interference. \textit{IEEE Transactions on Signal Processing}, \textbf{55}(6), 2386-2394.\\

Fortunati, S., Gini, F., Greco, M. S. (2016). The Misspecified Cramér-Rao Bound and its Application to the Scatter Matrix estimation in Complex Elliptically Symmetric distributions. \textit{IEEE Transactions on Signal Processing}, \textbf{64}(9), 2387-2399.\\

Huber, P. J. (1967). The behavior of Maximum Likelihood Estimates under Nonstandard Conditions. \textit{Proc. of the Fifth Berkeley Symposium in Mathematical Statistics and Probability}. Berkley: University of California Press.\\

Kay, S. (1998) \textit{Fundamentals of Statistical Signal Processing: Detection Theory}, Englewood Cliffs, NJ, USA: Prentice-Hall.\\

Kano, Y., Berkane, M, Bentler, P. M. (1993). Statisitcal inference based on Pseudo-Maximum Likelihood Estimators in elliptical populations. \textit{Journal of the American Statistical Association}. \textbf{88}(421), 135-143.\\

Kelly, E. J. (1986). An Adaptive Detection Algorithm.  \textit{IEEE Transactions on Aerospace and Electronic Systems}. \textbf{22}(1), 115-127.\\

Kent, J. T. (1982). Robust Properties of Likelihood Ratio Test. \textit{Biometrika}. \textbf{69}(1), 19-27.\\

Ollila, E., Tyler D. E., Koivunen, V., Poor, H., V. (2012). Complex elliptically symmetric distributions: survey, new results and applications. \textit{IEEE Transactions on Signal Processing}. \textbf{60}(11), 5597-5625.\\

Robey, F. C., Fuhrmann, D. R., Kelly, E. J., Nitzberg, R. (1992). A CFAR adaptive matched filter detector. \textit{IEEE Transactions on Aerospace and Electronic Systems}. \textbf{28}(1), 208-216.\\

White, H. (1982). Maximum likelihood estimation of misspecified models. \textit{Econometrica}. \textbf{50}, 1-25.\\

\end{document}